\documentclass[10pt, twocolumn, floatfix, superscriptaddress, aps]{revtex4-2}

\usepackage{amsmath, amssymb, graphicx, hyperref, xcolor, colortbl, braket}
\usepackage{cleveref}
\usepackage{comment}
\usepackage{mathtools}
\newcommand{\addt}[1]{\textcolor{black}{#1}} %
 
\begin{document}
\title{Origin and stability of generalized Wigner crystallinity in triangular moiré systems} 
\author{Aman Kumar}
\email{akumar@magnet.fsu.edu}
\affiliation{National High Magnetic Field Laboratory, Tallahassee, Florida 32310, USA}
\affiliation{Department of Physics, Florida State University, Tallahassee, Florida 32306, USA}

\author{Cyprian Lewandowski}
\email{clewandowski@magnet.fsu.edu}
\affiliation{National High Magnetic Field Laboratory, Tallahassee, Florida 32310, USA}
\affiliation{Department of Physics, Florida State University, Tallahassee, Florida 32306, USA}

\author{Hitesh J. Changlani}
\email{hchanglani@magnet.fsu.edu}
\affiliation{National High Magnetic Field Laboratory, Tallahassee, Florida 32310, USA}
\affiliation{Department of Physics, Florida State University, Tallahassee, Florida 32306, USA}
\date{\today}

\begin{abstract}
\noindent
\MakeUppercase{\textbf{Abstract}}\\
\addt{Generalized Wigner crystals (GWC) on triangular moiré superlattices, formed from stacking two layers of transition metal chalcogenides, have been observed at multiple fractional fillings [Nature 587, 214 - 218 (2020), Nat. Phys. 17, 715 - 719 (2021), Nature 597, 650 - 654 (2021)]. Motivated by these experiments, tied with the need for accurate microscopic descriptions of these materials, we explore the origins of GWC at $n=1/3$ and $2/3$ filling. We demonstrate the general limitations of theoretical descriptions relying on finite-range, versus long-range interactions, however, we clarify why some properties are captured by an effective nearest-neighbor model. We study both classical and quantum effects at zero and finite temperatures, discussing the role of charge frustration, identifying a ``pinball" phase, a partially quantum melted GWC, with no classical analog.  Our work addresses several experimental observations and makes predictions for how many of the theoretical findings can be potentially realized in future experiments.}
\end{abstract}
{\maketitle}

\noindent
\MakeUppercase{\textbf{Introduction}}
\\
The study of moiré systems has opened up multiple new avenues of research in condensed matter physics~\cite{Andrei2021,Mak2022,Balents2020}. Previously conjectured phases of electronic matter that were attributed to just the imagination of creative theorists are now being realized and observed in these platforms, e.g., see refs.~\cite{cao2018n, tang2020n, mak_xu2020correlated, feng_li2021imaging,Kang2024,Park2023}. The unique properties of moiré systems that allow for these proposed phases to be readily realized rely on the significant reduction in overall energy scales enabled by the moiré superlattice (while keeping strong correlations intact) combined with the tunability of carrier densities by the electrostatic gating characteristic of 2D materials. An excellent example of such theoretical and experimental synergy is shown by recent works that demonstrate how \addt{the} triangular lattice Hubbard model qualitatively explains the properties of moiré transition metal dichalcogenide systems (TMD)~\cite{ mak_xu2020correlated, mak_xu2020correlated, Jin2021, feng_li2021imaging, Regan2020, Huang2021, pan2020quantum,Wu_Macdonald,PhysRevB.107.235131,PhysRevB.108.245113} that display itinerant ferromagnetism (above half filling)~\cite{Shastry_Anderson, Davydova_Fu, Lee_Sharma_Vafek_Changlani_2023} and antiferromagnetism from kinetic frustration (below half filling)~\cite{Haerter_Shastry_2005, Haerter_Peterson_Shastry_2006}, with observable effects at ``intermediate" temperatures~\cite{Lee_Sharma_Vafek_Changlani_2023, Morera_2023, Xu_Greiner_2022}. 
\begin{figure}
\includegraphics[width=0.98\columnwidth]{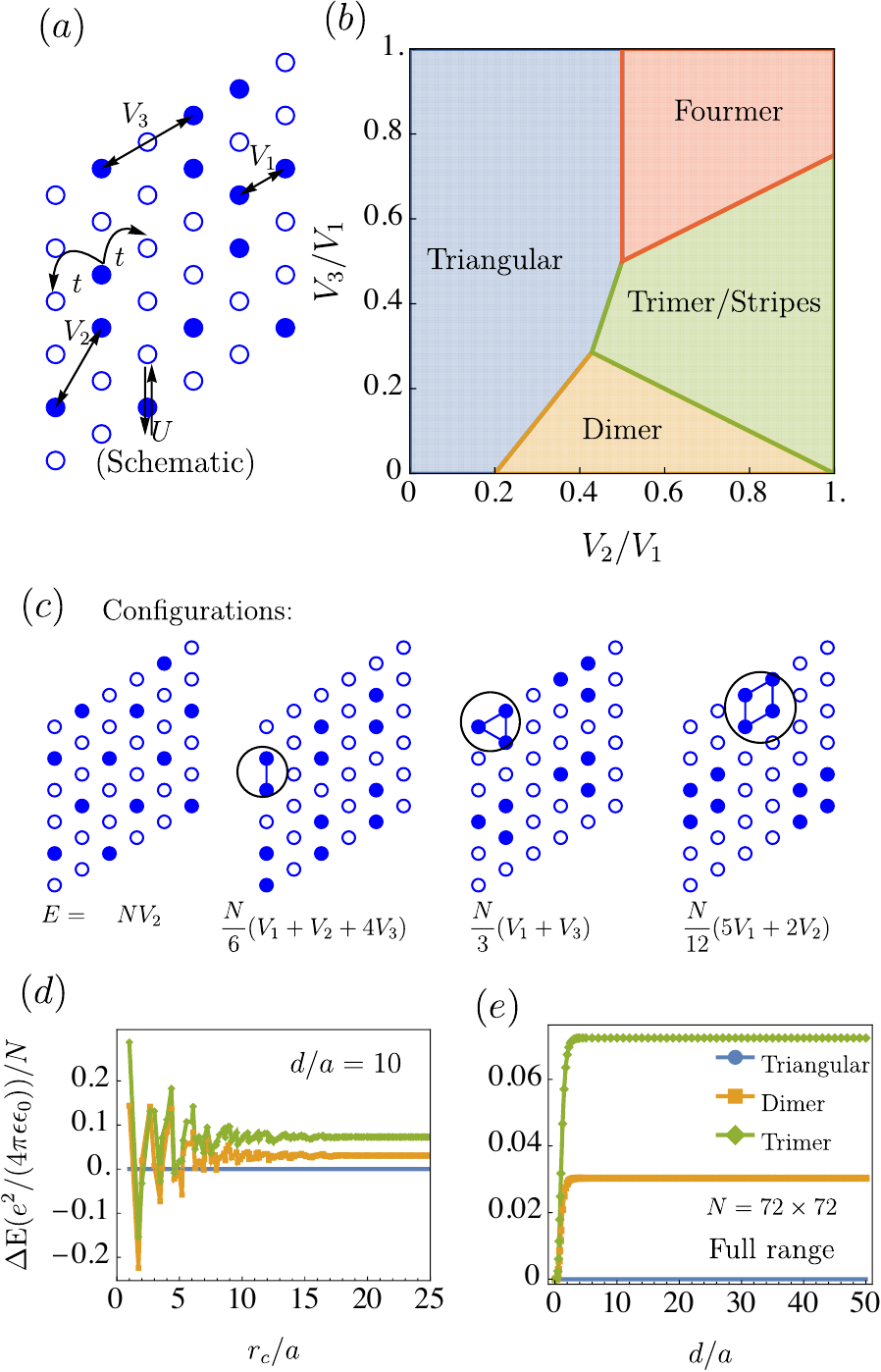}
\caption{\addt{ \bf{Hamiltonian, classical phases, and energetics of the extended Hubbard model on the triangular lattice.}}\label{fig:schematic} (a) A schematic of \addt{the extended Hubbard model as in Eq.~\eqref{eq:model_ham}, with nearest neighbor hopping $t$, on-site Hubbard interaction $U$, and nearest, next-nearest and next-next-nearest neighbor interactions denoted by $V_1,V_2,V_3$ respectively}. (b) \addt{Locations of lowest energy states in phase space, restricted to $n=1/3$ charge configurations shown in panel c, in the $V_1-V_2-V_3$ model ($V_1>0, U \to \infty$, $t=0$).}. 
(c) \addt{Charge} configurations for $n=1/3$ showing the triangular, dimer, trimer, fourmer crystals, \addt{from left to right}, and expressions for the corresponding total energy for $N$ sites. (d, e) The \addt{individual } energy differences of the dimer and trimer configurations, \addt{each with respect to the triangular configuration, as a function of interaction cutoff $r_c/a$ for a fixed $d/a=10$ (panel d), and gate distance $d/a$ for $r_c \rightarrow \infty$ (panel e). The calculation is carried out on a $N=72\times72$ system with periodic boundary conditions. The y-axis label for panels d,e is the same}.
}
\label{fig:classical}
\end{figure}
While the Hubbard model description of the moiré TMD systems has undoubtedly been very successful \cite{Tang2020, Wu_Macdonald}, a crucial aspect missing 
is the role the long-range nature of the Coulomb interaction can play controlling the electronic phase diagram (e.g. see the discussion on the importance of non-local/finite range interaction terms in Hubbard model in refs. \cite{PhysRevLett.128.217202, Tan_2023,Camjayi2008}). In this work, we focus on the role of long-range Coulomb interaction in these systems and analyze the consequences of projecting 
it to a finite-range \addt{extended} Hubbard model. 
Specifically, we systematically explore, at zero and finite temperature, both finite-range (FR) and long-range (LR) Hubbard models given by,
\begin{equation}
\label{eq:model_ham}
 H = - \sum_{i<j, \sigma} t_{ij} c^{\dagger}_{i,\sigma} c_{j,\sigma} + \textrm{h.c.} + U \sum_{i} n_{i,\uparrow} n_{i,\downarrow} + \sum_{i<j} V_{ij} n_i n_j,
\end{equation}
where $i,j$ refer to the sites of a (moiré) triangular lattice, $\sigma$ is the spin index, and $t_{ij}$, $U$ and $V_{ij}$ refer to the hopping, on-site \addt{Hubbard} interaction strength and density-density interactions respectively, see Fig.\ref{fig:classical}a. \addt{$c^{\dagger}_{i,\sigma}$ and $c_{i,\sigma}$ are fermion creation and annihilation operators respectively} and $n_{i,\sigma} \equiv c^{\dagger}_{i,\sigma}c_{i,\sigma}$ and $n_i \equiv n_{i,\uparrow} + n_{i,\downarrow}$ are number operators. We focus specifically on the fillings ($n$) of $1/3$ and $2/3$ to make the physics of \addt{the LR} Coulomb interaction truncation the center of our discussion, leaving an exhaustive study of other fillings for future work. 

The experimental platform that realizes the relevant physics for this manuscript is the moiré heterobilayer $\mathrm{WSe}_2$/$\mathrm{WS}_2$ system \cite{mak_xu2020correlated, Jin2021, feng_li2021imaging, Regan2020, Huang2021}, which hosts generalized Wigner crystals (GWC) at fillings commensurate with the moiré superlattice enabled by the $\mathrm{WSe}_2$,$\mathrm{WS}_2$ lattice mismatch \cite{Wu_Macdonald, PhysRevB.102.201115}. Interest in GWCs has been fueled by recent experiments that used optical probes \cite{mak_xu2020correlated} and scanning tunneling microscopy (STM) \cite{feng_li2021imaging} to confirm their presence in moiré TMDs. In the STM work in particular, a state consistent with an expected $\sqrt{3} \times \sqrt{3}$ \addt{triangular charge} order was found in small local patches, however, vacancies in the underlying charge order with motifs that visually resemble other orders are also seen. The nature of these insulating GWC orders has been studied theoretically through a \addt{combination} of theoretical techniques starting from a continuum-model or Hubbard-model-like description and employing field theory, Hartree-Fock, \addt{dynamical mean field theory}, Monte-Carlo, density matrix renormalization group (DMRG), \addt{and} exact diagonalization (ED)~\cite{Tan_2023,PhysRevB.106.155145,PhysRevB.103.155142, PhysRevLett.132.076503, Matty2022, PhysRevB.103.L241110,PhysRevB.107.235131, Padhi_2021,arxiv.2409.11202}. 
\addt{We note that the term GWC is often used interchangeably with charge density wave (CDW), however, the difference is that CDW emerges as an instability due to Fermi surface nesting whereas in a GWC the charge localization is interaction driven and the kinetic energy is strongly quenched.}

Our work addresses currently unresolved aspects of these experiments, contributing to the theoretical understanding of the moiré TMD systems and presenting both zero-temperature and finite-temperature studies in a unified framework. Our paper is structured in three parts, each addressing one precise objective. The first objective is to demonstrate that while FR Hubbard models have the essential elements to describe a GWC, they \addt{are typically} inadequate for a quantitative explanation of the experiments if the \addt{LR tail is rigidly truncated}. This analysis involves a general exploration of \addt{energetically competitive phases at} the classical ($t=0$, $U \rightarrow \infty$) and quantum mechanical level for the Hamiltonian in Eq. \eqref{eq:model_ham} at zero temperature. 
The next objective is to understand the finite temperature melting of these GWCs. We build on the classical treatment that accurately modeled many aspects of the experiments~\cite{mak_xu2020correlated}, but did not address the role of quantum \addt{effects} - after all, it is not \textit{a priori} clear if the melting of the GWC (at $\sim 37$ K) has any quantum origin given that the kinetic energy is $t \sim 20$ K. Our numerical calculations explain the small but experimentally detectable \cite{mak_xu2020correlated} (few Kelvin) difference in the transition temperatures of the $n=1/3$ and $2/3$ GWC, that are exactly dual to one another at the classical level. 
A natural outcome of the first two objectives is the partial reconciliation of conflicting parameter sets chosen by different groups~\cite{mak_xu2020correlated, pan2020quantum, Motruk_2023}.
Our third objective is to predict the extent of the stability of GWC with screening gate distance (see also ref.~\cite{PhysRevB.107.235131}) - a key component of the moiré TMD experimental setup. This objective aims to pave the way for informed experiment design by determining at what temperatures the magnetic correlations present in GWC \cite{PhysRevB.104.214403, pan2020quantum,PhysRevLett.127.096802,PhysRevB.108.245113, PhysRevB.105.041109,Kaushal2022}, are most prominent and thus most easily detectable. 
\\
\\
\MakeUppercase{\textbf{Results}}\\
\\
{\bf{Classical phase diagram and competing charge-ordered states}}
\\
To develop an intuition for the GWC problem, we first construct the classical ground state phase-competition diagram ($t=0$ and $U \rightarrow \infty$) for $n=1/3$ of the Hamiltonian in Eq. \eqref{eq:model_ham} by \addt{exactly enumerating all states on a $6\times$6 lattice with periodic boundary conditions and picking the ground state(s).} This calculation serves as a helpful starting point for furhter analyses as the leading energy scale of the problem is the Coulomb interaction between electrons trapped in deep moiré potential wells\cite{PhysRevLett.121.026402} - a point to which we return later. We consider nearest (NN, strength $V_1=1$), next-nearest (\addt{NNN}, strength $V_2$) and next-to-next nearest neighbor interactions (\addt{NNNN}, strength $V_3$); our results are summarized in Fig.~\ref{fig:classical}b,c. No double occupancy is allowed, i.e., $U \to \infty$, and spin is irrelevant. The $n=2/3$ results can be obtained by swapping the roles of particles and holes. \addt{(For systems larger than $6 \times 6$, we have carried out additional low temperature classical Monte Carlo calculations to navigate the low energy landscape. We have observed the existence of regions of phase space where the configurations reported in Fig.~\ref{fig:classical}b and c do not strictly correspond to the true ground state, just low energy states.)}
\addt{
\begin{table}
\centering
\begin{tabular}{ c|c|c|c } 
 \hline 
 & \multicolumn{3}{c}{Parameter values} \\
 \hline
Model & {$t (\mathrm{meV})$}  & $U (\mathrm{meV})$ & $V_{ij} (\mathrm{meV})$ \\
 \hline                        
 1 & 1.81 &    75 $t$ &  $V_1$ = 10.5 $t$ \cite{Motruk_2023}\\
 2 &  0   &   $\infty$ &  $a=7.98$ nm, $\epsilon = 3.9$, $d/a=10$ \cite{mak_xu2020correlated} \\ 
 3 &  1.81   &   75 $t$ &  $a=7.98$ nm, $\epsilon = 3.9$, $d/a=10$ \\
 4 & 1.81 &   75 $t$ & $V_1$=39.92, $V_2$= 20.44, $V_3$=16.89 \\       
\hline
\end{tabular}
\caption{\addt{Summary of parameter sets employed in this work, which we refer to as models 1-4. All energy units are in meV.
$a$ is the moiré lattice constant. 
For the double gated potential, $d$ is the separation between the gates. The functional form of this potential is stated in Eq.~\eqref{eq:coulomb_long_range} and its implementation is discussed in \addt{Supplementary Note 1}. Model 1 (ref.~\cite{Motruk_2023}) and model 2 (ref.~\cite{mak_xu2020correlated}) 
appeared previously in the literature. ref.~\cite{Motruk_2023} also considered non-zero further range hoppings ($t_2,t_3$), which we have ignored in this work. Model 4 is obtained by truncating model 3 to next-next neighbor interactions.}}
\label{table:moiré_parameters}
\end{table}
}


For small $V_2, V_3$ the $\sqrt{3} \times \sqrt{3}$ \addt{triangular} GWC is stable, as the total charge configuration completely avoids the $V_1$ and $V_3$ costs at the expense of a $V_2$ cost. However, other phases are favored for modest $V_2/V_1$ and $V_3/V_1$. For example, for $V_3 = 0 $, \addt{  $ 0.2 \le V_2/V_1 \le 1$} stabilizes a ``dimer" crystal, the state that arises out of a compromise - the system pays some $V_1$, $V_2$ and $V_3$ costs by pairing up into dimers, as shown in Fig.~\ref{fig:classical}c. Interestingly, \addt{on the $6 \times 6$ torus, this arrangement is found to be degenerate with 3707 other states (degeneracies include those beyond those arising from translation and different orientations of the dimer crystal) suggesting a high amount of ``charge frustration". This degeneracy persists for $V_3=0$ for all $V_2/V_1$ as long as the system is in the dimer phase. (On larger system sizes, we observe a finite residual entropy per site from integration of the specific heat from classical Monte Carlo, suggesting this number is exponential with system size). On adding small, but finite, $V_3 > 0$, the dimer (and symmetry related states) are favored, and this large degeneracy is lifted.}

\begin{figure}
\includegraphics[width=\linewidth]{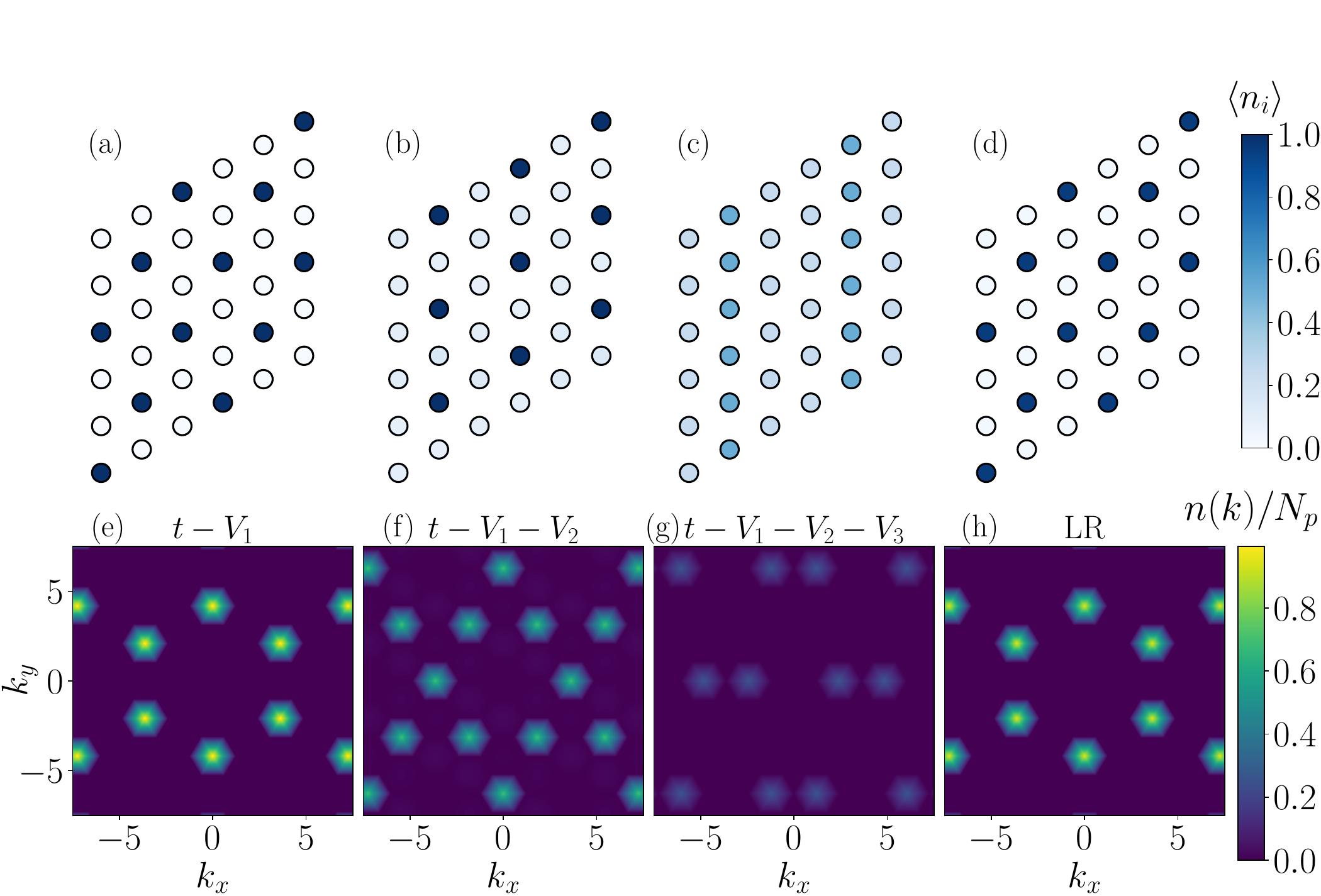}
\caption{\addt{ {\bf{Charge ordered quantum phases for a series of extended Hubbard models.}} (a-d) show the expectation value of the local charge density $\langle n_i\rangle$ on every site $i$ of a $6 \times 6$ torus for $n=1/3$, computed in the quantum ground state (obtained from DMRG) for a series of models. These models were obtained by truncating the LR model (model 3 in Table~\ref{table:moiré_parameters}) to various ranges (a) model 4 but with $V_2=V_3=0$ (b) model 4 with $V_3=0$ (c) model 4 (d) model 3. The $y$ axis is vertical and the $x$ axis is horizontal. (e-h) show the corresponding absolute value of the Fourier transform of the charge distribution in momentum space; $n(\vec{k})$ is defined as  $\Big|\sum_i \left( \langle n_i \rangle - \frac{1}{3} \right) e^{i \mathbf{k} \cdot \mathbf{r_i}}\Big|
$ and $N_p$ is the number of particles (i.e. $N/3$). The DMRG simulations utilized a maximum bond dimension of up to 10000 and were carried out in the $S_z=0$ sector.}} 
\label{fig:quantum}
\end{figure}

At \addt{much} larger $V_3$, a crystal is formed out of a cluster of three charges, which we refer to as a ``trimer". This state pays no $V_2$ cost but pays some $V_3$ costs between trimers. It is exactly degenerate with a period-3 ``stripe" state, a periodic arrangement of one-dimensional lines of charges. In the case of this state, which will feature in Fig. \ref{fig:quantum}, all $V_1$ and $V_3$ costs are paid within each one-dimensional stripe. \addt{(On larger sizes we have observed the existence of lower energy amorphous states, featuring short stripes and trimers.)} Then, when $V_2$ and $V_3$ are both large, a ``fourmer" crystal, shown in Fig.~\ref{fig:classical}c, is stabilized. In general, we note that at low density the system avoids energy penalties by clustering particles, even in the absence of any attractive interactions - reminiscent of a pairing mechanism discussed in ref \cite{Crpel2021} and recent reports of ``bubble phases" in Landau levels \cite{PhysRevLett.131.226501}.

We now ask where in the phase-competition diagram of Fig. \ref{fig:classical}b are 
typical moiré TMD materials expected to be. To answer that question, it is necessary to compute the extended $V_{ij}$ Hubbard parameters. Following ref.~\cite{mak_xu2020correlated}, we consider a double-gate screened Coulomb interaction 
\begin{equation}
\label{eq:coulomb_long_range}
V(\vec{r}) = \frac{e^2}{4\pi \epsilon \epsilon_0 a }\sum^{k=\infty}_{k=-\infty} \frac{(-1)^k}{\sqrt{\Big(\frac{kd}{a} \Big)^2 + \Big( \frac{|\vec{r}|}{a} \Big) ^2}}, 
\end{equation}
\addt{where $a$ is the moiré lattice constant, $d$ is the separation between gates, $\epsilon$ is the dimensionless dielectric constant, $\epsilon_0$ is the vacuum permittivity, and $\vec{r}$ refers to the vector connecting the two moiré lattice sites}. The above form of the Coulomb interaction assumes two gates symmetrically located (above and below) at a distance $d/2$ from the moiré system, as relevant for the experiment of ref.~\cite{mak_xu2020correlated}, however our conclusions should generalize, with appropriate quantitative modifications, to other gate configurations~\cite{feng_li2021imaging}. In the above calculation of \addt{$V_{ij} \equiv V(\vec{r}_i - \vec{r}_j)$} parameters, we also assume that the Wannier functions are localized on a length scale much smaller than the moiré lattice constant (which is expected for this topologically trivial system \cite{Wu_Macdonald, PhysRevB.102.201115}). In such a case it is reasonable to approximate the interactions as those between point charges -- \addt{we employ this approximation throughout this work.} We clarify that, in principle, the Wannier function overlap can be incorporated (e.g. see ref. \cite{Wu_Macdonald, pan2020quantum, PhysRevLett.128.217202}) into estimating the on-site Hubbard $U$ and extended $V_{ij}$. 
While this treatment is expected to only provide a small correction to the extended $V_{ij}$ parameters, it is essential for the estimation of the Hubbard $U$ e.g., as discussed in ref.~\cite{Motruk_2023}.
However, since our work focuses on fillings away from one electron per moiré cell, 
\addt{our results are not significantly influenced by the precise value of $U$}- all that is important for our conclusions is that $U$ is large enough to prevent any appreciable double occupancy on a site. 

As is explained in \addt{Supplementary Note 1}, we treat the \addt{LR} nature of the interaction by tiling the finite size cluster (the fundamental unit cell) multiple times to cover all of space, and then use the arrangement to calculate the effective interaction between any two sites (including the site with itself) in the fundamental unit cell. 
\addt{This effective Hamiltonian is simulated with classical and quantum methods. We note that the exponential decay of the potential with distance guarantees rapid convergence of the procedure with increasing the number of tiles. We also clarify that these effective interactions must not be confused with the renormalization of the NN interaction due to the LR tails that we discuss later in the paper.} 

To highlight how the competition between ground states is sensitive to the range of truncation, we consider a sequence of models with increasing range $r_c$ (the distance at which the interaction is truncated) 
i.e., increasing number of non-zero $V_{ij}$ extended Hubbard model terms in Eq.\eqref{eq:model_ham}. Specifically, we take $V(\vec{r})$ to have the form in Eq.\eqref{eq:coulomb_long_range} for $ r \equiv |\vec{r}| \leq r_c$ and $V(\vec{r}) = 0$ for $ r > r_c$. In Fig. \ref{fig:classical}d, we plot, for $d/a=10$, \addt{the} energy difference (per site) of each of the competing states  we found in Fig. \ref{fig:classical}b \addt{with respect to the energy of the triangular crystal}. 
For NN interactions, indeed, the triangular 
$\sqrt{3} \times \sqrt{3}$ crystal is the ground state, however, truncating to NNN yields the dimer crystal and truncating to higher neighbor interactions generically gives more complex ground states. Fig. \ref{fig:classical}e shows the competition between ground states for the LR interaction (i.e., $r_c \to \infty$) as a function of $d/a$. When LR interaction is considered, we find that the $\sqrt{3} \times \sqrt{3}$ GWC is always the ground state for any gate distances $d/a \gtrsim 1$.
\\
\noindent
\\
{\bf{Quantum treatment, ``pinball phase'', and state selection}}
\\
The small energy differences between competing classical states, particularly when the interactions are long-ranged, as in Fig.~\ref{fig:classical}d, highlight the need to account for quantum effects accurately. To do so, we consider the simplest possible kinetic term, the NN hopping, which was estimated to be much larger than other hoppings (see Table ~\ref{table:moiré_parameters}). In principle, in analogy to the $V_{ij}$ Hubbard terms in Eq.\eqref{eq:model_ham}, the kinetic term could have contributions beyond NN hopping. However, since kinetic energy is significantly suppressed in GWCs and largely acts as a perturbation to ground states selected by the Coulomb interactions, we argue that it is sufficient to consider just the \addt{dominant} NN term to capture the relevant charge physics we discuss in this work. In future work, it would be interesting to study the role of extended hoppings, especially in parts of the phase diagram where multiple charge or magnetic orders compete closely~\cite{Motruk_2023, pan2020quantum}.

We obtain the quantum mechanical ground state on finite clusters using matrix product state (MPS) based DMRG calculations \cite{DMRG_White,tenpy}. \addt{We work with fixed particle number and total $S_z=0$}. On a given system size, DMRG is limited only by finite bond dimension, when this is inadequate DMRG favors low entanglement states such as those with broken translational symmetry. Given the 1-D nature of the MPS which is used to ``snake through" a 2-D system, the accuracy of DMRG is typically limited on systems with periodic boundaries (torus in 2D) or those with long correlation lengths~\cite{Stoudenmire_White_review_2012}. On the other hand, the torus geometry has the advantage of having no boundary effects and so is possibly more representative of the thermodynamic limit. Thus we interpret the results of our bond dimension-limited runs on tori, that favor symmetry-broken states with low entanglement, to be genuinely representative of the underlying physics, despite them not being exact eigenstates. To supplement these results and build confidence in our results, we have carried out additional DMRG calculations on cylinders. The quasi-1D nature of the cylinders, while conducive for DMRG, is known to affect the physics especially if the length of the cylinder is taken to be much larger than its width~\cite{Sandvik_cylinders_2012}. \addt{Despite} these caveats, we arrive at similar qualitative conclusions on cylinders, see additional results and discussion in \addt{Supplementary Note 2}.

The representative results for $n=1/3$ for a sequence of FR and LR Hubbard models on a $6\times 6$ torus (computed with a maximum bond dimension of 10000) are shown in Fig.~\ref{fig:quantum}a-d. All the finite-range models discussed in this section \addt{are obtained by combining the 
parameters from ref.~\cite{mak_xu2020correlated} and ref.~\cite{Motruk_2023} (i.e. model 3 with $\epsilon = 3.9$ and $d/a=10$, see Table~\ref{table:moiré_parameters}) and then setting all $V_{ij}$ beyond a certain range to zero. (Note that when model 3 is truncated to $V_1, V_2,V_3$ it yields model 4.)} We plot the charge density ($\langle n_i \rangle$) on every lattice site and also show the Fourier transform of the density correlation function, Fig.~\ref{fig:quantum}e-h. 


\begin{figure*}
\includegraphics[width=\textwidth]{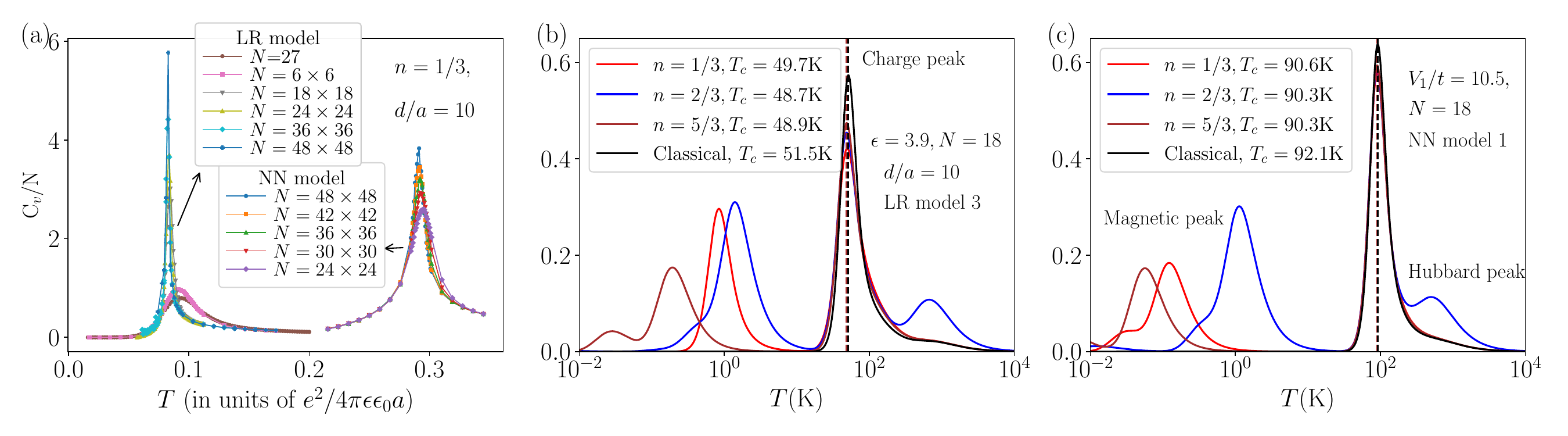}
\caption{\addt{ {\bf Specific heat per site ($C_v/N$) as a function of temperature $(T)$, for various models considered in this work.} (a) $C_v(T)/N$ for the LR model (model 2 in Table~\ref{table:moiré_parameters}) and the NN model (model 2 but truncated to have only $V_1$ non-zero) computed with classical Metropolis Monte Carlo for multiple system sizes. The classical model with $t=0$, $U \rightarrow \infty$ is particle-hole symmetric, so the results for $n=1/3$ and $n=2/3$ are identical. (b,c) show $C_v (T)/N$ for the quantum models obtained from exact diagonalization for $N=18$ sites (see \addt{Supplementary Note 3}) for three densities $n=1/3,2/3$ and $5/3$. (b) $C_v(T)/N$ for the LR quantum model (model 3) and (c) $C_v(T)/N$ for a NN (model 1). 
Additionally, in panels (b) and (c) we show the ``classical" result for $n=1/3$ for the same system size, obtained by setting $t=0$ but not changing the other parameters (i.e. keeping $U$ large, but finite).
}}
\label{fig:specific_heat_classical_quantum}
\end{figure*}

In agreement with expectations, we find \addt{that} for the NN model ($V_1/t = 22.1$) the $\sqrt{3} \times \sqrt{3}$ \addt{triangular crystal seen classically is stable to the introduction of a hopping i.e. quantum effects}, see Fig.~\ref{fig:quantum}a. The charge density is sharply localized on the sites of the $\sqrt{3} \times \sqrt{3}$ crystal, c.f. Fourier transform in Fig.~\ref{fig:quantum}e, confirming the identification of this charge density wave ground state with that of a GWC.

However, the ground state of the NNN model \addt{($t-V_1-V_2$ model, i.e. model 4 with $V_3=0$)} is strikingly distinct from the classical expectation - we find no evidence of stabilization of a dimer crystal or any of the other classically degenerate ground states. Instead, we see the appearance of charge centers (with $\langle n_i\rangle \approx 1$) at sites of a triangular crystal with a spacing of two (moiré) lattice constants, Fig. \ref{fig:quantum}b,f. This emergent $2 \times 2$ triangular crystal accounts only for 1/4 filling, and the remaining $1/3 - 1/4 = 1/12$ charge density was found to be delocalized on the other sites. This ``partially melted" GWC arises out of a compromise - the system avoids paying both $V_1$ and $V_2$ costs between the charges on the ``pinned" sites and also benefits from kinetic energy delocalization of the remaining charges. There is also a Coulomb energy cost associated with the delocalized charges interacting with the pinned charges, but overall, the first two effects dominate, leading to the stabilization of such a crystal. We envisage that while this \addt{NNN} model does not govern the realistic moiré TMD system, the phase we have found here could be potentially realized in other platforms such as cold atoms~\cite{Xu_Greiner_2022,RevModPhys.91.015005}. 
\addt{We also note that previous theoretical work established a ``pinball liquid phase" \cite{Hotta_2006, Fratini_pinball} at a different density $n = 1/2$ (quarter filling) and in an extended Hubbard model on the triangular lattice with only $V_1$ interactions i.e. $V_2=0$. For $n=1/2$ the pins are arranged on a $\sqrt{3} \times \sqrt{3}$ triangular crystal (in contrast to the $2\times 2$ cell seen here), but the essential physics of coexisting insulating and delocalized degrees of freedom appears to be similar.}

\addt{The NNNN model ($t-V_1-V_2-V_3$ model for the parameter set in Table~\ref{table:moiré_parameters} i.e. model 4)} yields a period-3 stripe ground state, resembling its classical counterpart - see Fig. \ref{fig:quantum}c. Here, one of three equivalent directions of the triangular lattice is spontaneously selected. Unlike the classical result, however, charge localization is weakened in strength due to quantum \addt{effects}, leading to smaller peaks in the Fourier transform of the charge density profile, Fig. \ref{fig:quantum}g. It is interesting to observe that extended stripes are favored over local trimers - a possible mechanism is ``order by disorder" which dictates how quantum state selection occurs among a collection of classically degenerate states~\cite{Villain_1980, Henley_obd_1989} and which has also been shown to be relevant for other materials with charge order~\cite{Subires2023}.

Most importantly, we find that the LR model \addt{(i.e. model 3)} stabilizes a $\sqrt{3}\times\sqrt{3}$ \addt{triangular} GWC, see Fig. \ref{fig:quantum}d. \addt{The} resulting charge density localization is identical to 
the \addt{NN} model, with only minor quantitative differences in the densities on the GWC sites, c.f. Fig. \ref{fig:quantum}e,h.
This result suggests that the NN model, with appropriate renormalization of the extended interaction $V_1$, may capture the essential physics of GWC melting at finite temperature, which we will explore next.
\\
\\

{\bf{Finite temperature properties of GWC}}
\\
We now proceed to analyze the finite temperature properties of the GWCs. We begin by considering the classical case first ($t = 0$, $U\to \infty$ in Eq.\eqref{eq:model_ham}). Fig.~\ref{fig:specific_heat_classical_quantum}a shows the specific heat per site $C_v/N$ \addt{as a function of temperature} for the LR and truncated NN classical models for the case of $d/a=10$ \addt{i.e. model 2, and model 2 with only $V_1 \ne 0$, respectively. Calculations were done with classical Metropolis Monte Carlo for $n=1/3$ for a range of system sizes; the $n=2/3$ results are identical due to particle-hole symmetry. While finite size effects are naturally expected, we observe that the location of $T_c$ for system sizes as small as $N=27$ sites is broadly consistent with much larger sizes (to within 10\%).}
Importantly, we find that despite \addt{both models} having the same classical ground state, the \addt{charge ordering temperature} $T_c$ of the LR model is a factor of $\approx 3.6$ smaller than that of the NN model;  indicating that the long-range tail significantly renormalizes the NN interaction. \addt{We will revisit this finding soon and also provide an explanation for it.}

With \addt{an expectation of the classical melting} behavior in place, we now consider a quantum mechanical finite-temperature study of the Hamiltonian in Eq. \ref{eq:model_ham} \addt{, specifically for model 1 and 3 in Table~\ref{table:moiré_parameters}, using ED (full diagonalization) and the finite temperature Lanczos method (FTLM)~\cite{Jaklic_Prelovsek, Prelovsek_Bonca}, see the Methods section}. 
\addt{Since spin has an important role to play in the quantum simulations, we have also shown results for the $n=5/3$ case; this density is a spinful particle-hole partner of the $n=1/3$ case, its Hamiltonian is equivalent to that of $n=1/3$, but with $t \rightarrow -t$.} 

Given that meaningful conclusions and trends can be drawn from small clusters, we compute the quantum mechanical specific heat for the $N=18$ system (see \addt{Supplementary Note 3} for cluster shape and symmetries used \addt{and Supplementary Note 4 for simulations on more sizes}). We also consider (using the same \addt{ED} procedure) the ``classical" case with $t=0$ but with otherwise identical parameters, i.e., $U$ large but finite. Our results are shown in Fig.~\ref{fig:specific_heat_classical_quantum}b,c on a log temperature scale for the LR \addt{(model 3)} and NN \addt{(model 1)} case, respectively. The NN interaction strength \addt{in model 1} is in the ballpark of that studied in ref.~\cite{Motruk_2023} where $\epsilon$ was chosen to be $10$. 

The differences in the choices of $\epsilon$ (see for example work of ref.~\cite{mak_xu2020correlated} and ref.~\cite{Motruk_2023}) followed by truncation (or lack of it) can potentially reconcile parameter sets that otherwise look very different. 
In particular, two effects compete with one another: increasing $\epsilon$ has the effect of decreasing the overall strength of Coulomb interactions and hence lowering the ordering temperature, and truncating the LR interaction eliminates the overall \addt{renormalization} that comes from the \addt{LR} tail, in turn increasing the ordering temperature. The end result is that the effective $V_1/t$ used in the NN model is approximately $10.5$~\cite{Motruk_2023}. We find that this, in turn, results in the GWC temperature melting temperature that is a factor of roughly two higher (see Fig.~\ref{fig:specific_heat_classical_quantum}c) than that measured in experiment and reproduced by classical Monte Carlo with the full LR model with $\epsilon = 3.9$~\cite{mak_xu2020correlated}.

The quantum-mechanical finite-temperature simulations show that the entropy is released in \addt{multiple} steps for both LR and NN models. \addt{ For example, for $n=2/3$ there are at least three prominent and distinct
temperature scales associated with either a crossover or transition. The highest scale 
is associated with the large Hubbard $U$, as has been noted previously in ref.~\cite{Lee_Sharma_Vafek_Changlani_2023} in the context of the on-site Hubbard model. The intermediate scale corresponds to the melting of the charge order and has been accessed experimentally~\cite{mak_xu2020correlated}. The low-temperature bump (crossover) corresponds to 
the destruction of finite-range magnetic correlations (there 
is no true long order at finite temperature due to the Mermin-Wagner theorem~\cite{mermin1966prl}).
For $n=1/3$ and $n=5/3,$ the Hubbard feature, while present, is greatly suppressed. The charge ordering scales are similar, but not exactly the same, and the magnetic scales are different from one another and the $n=2/3$ case.}
As magnetic correlations are quantum 
mechanical in nature, they are completely absent in the classical calculations. 
\addt{(We also note the presence of an additional ultra-low temperature feature which is expected to emerge from small but finite size gaps, and hence we ignore its presence for the discussion here.)} 

The suppression of kinetic energy in 
the GWC means that its melting is essentially, \addt{but not completely}, classical in origin. \addt{This is true for both the NN and LR models. For the NN model with 
$V_1/t = 10.5$ \addt{(model 1)}, we find a small difference (within $\approx$ 0.3 K) in the 
locations of $T_c$ for $n=1/3, 2/3$ and $5/3$, and an approximately $2$ K difference with the "classical" case where $n=1/3$ and $t=0$, while all other parameters were kept fixed (i.e. $U$ was kept finite).
We attribute these observations to the small amount of quantum melting of the ground states and the kinetic energy term on the triangular lattice breaking the particle-hole symmetry of the $t=0$ model.
} 
The LR model \addt{(model 3)} has effectively weaker 
\addt{NN} interactions \addt{which results in} a lower $T_c$.
\addt{Importantly, for this model the difference between the $n=1/3$ and $2/3$ cases is bigger i.e. approximately $1$ K, which is in the ballpark of the experimental findings~\cite{mak_xu2020correlated}.} 

\addt{As mentioned previously,} the difference between $n=1/3, 2/3, 5/3$ manifests itself even more prominently in the low-temperature magnetic features, which remain to be experimentally explored, c.f. Fig.\ref{fig:specific_heat_classical_quantum}b. Our calculations predict the magnetic crossover temperatures in the range approximately $0.2$ K - $1$ K. 
These magnetic crossover scales are \addt{potentially} within \addt{an} experimentally realizable range, suggesting that future experiments sensitive to spin texture (e.g., NV center scanning techniques, spin-polarized scanning tunneling microscopy or nanoSQUID) could resolve the nature of the various spin ground states\cite{PhysRevB.104.214403, pan2020quantum,PhysRevLett.127.096802,PhysRevB.108.245113, PhysRevB.105.041109,Kaushal2022}.

\begin{figure}
\includegraphics[width=1\columnwidth]{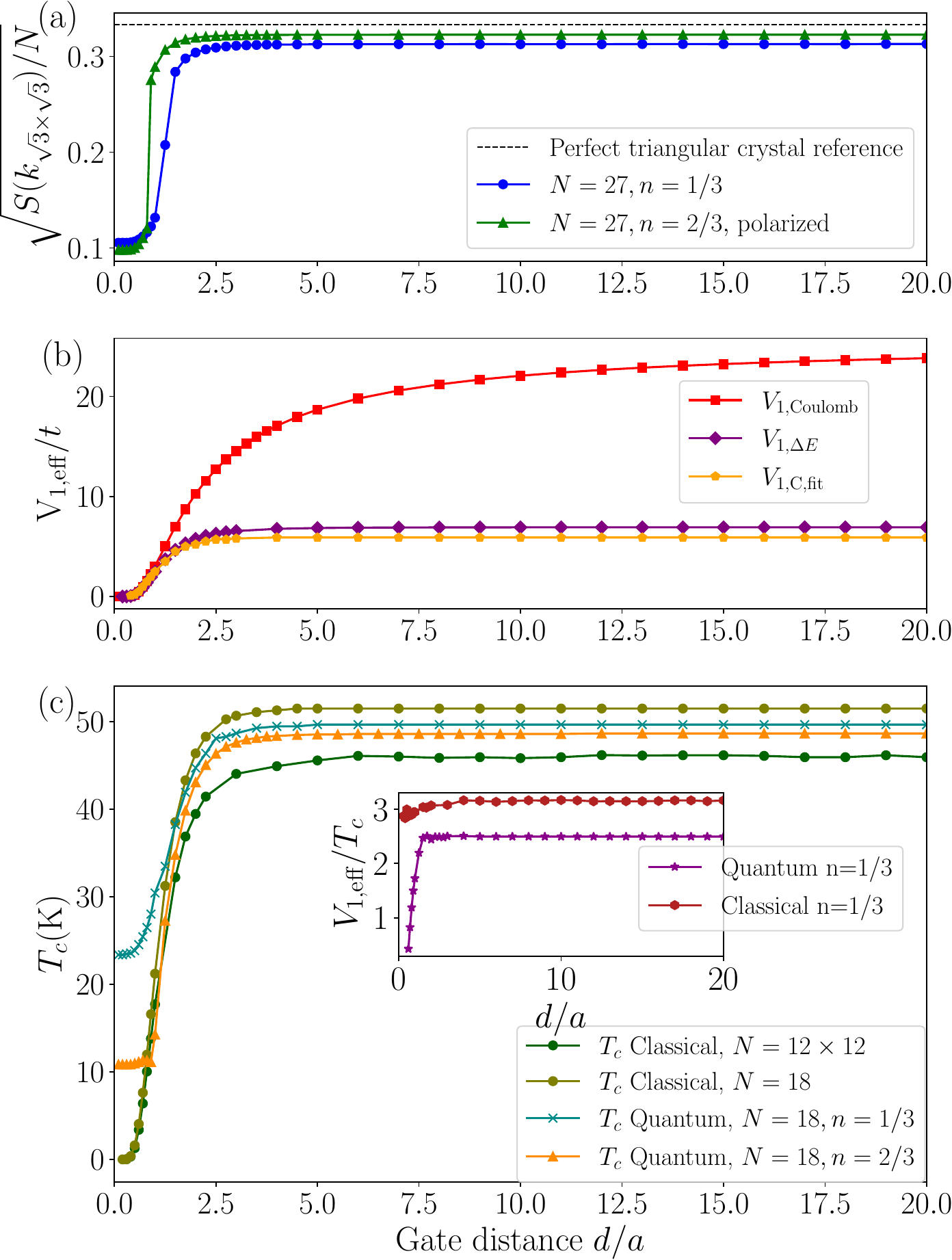}
\caption{\addt{ {\bf Dependence of effective ground state and finite temperature properties of GWCs on gate distance.} (a) Order parameter (defined in terms of the charge structure factor, $S( {\bf k}) \equiv \frac{1}{N} \sum_{i,j} \langle n_i n_j \rangle e^{i \mathbf{k} \cdot (\mathbf{r}_i - \mathbf{r}_j)}$ computed at a representative momentum point $k_{\sqrt{3} \times \sqrt{3}}$ that shows a peak for the $\sqrt{3}\times\sqrt{3}$ triangular charge order) in the quantum ground state, as a function of $d/a$. 
Calculations were carried out in the momentum (0,0) sector for $N=27$ and $n=1/3$ (in the $S_z=1/2$ sector) and $n=2/3$ (in the maximally polarized $S_z$ sector).
The parameters correspond to model 3 (Table~\ref{table:moiré_parameters}) for variable $d/a$. The dotted line represents the reference value for a perfect $\sqrt{3}\times\sqrt{3}$ triangular charge order. (b) Effective NN $V_{1, \text{eff}}/t$ interaction as a function of $d/a$ obtained from: the bare interaction potential ($V_{1, \text{Coulomb}} \equiv V(r=a)$, see Eq. \eqref{eq:coulomb_long_range}), one defect energy ($V_{1,\Delta E}$), and from fitting the specific heat curve of the LR model (model 3 with variable $d$) and NN model ($V_{1,\text{C,fit}}$). 
(c) The critical temperature $T_c$ corresponds to charge order melting transition for three cases: classical, quantum $n=1/3$, and $n=2/3$, as a function of $d/a.$ (Inset) Ratio of the $V_{1, \text{eff}}$ to $T_c$ using $V_{1,\Delta E}$ (classical) and $V_{1,\text{C,fit}}$ (quantum) from (b).}
}
\label{fig:delta_ED_Tc}
\end{figure}


\begin{figure}[h]
\includegraphics[width=1\columnwidth]{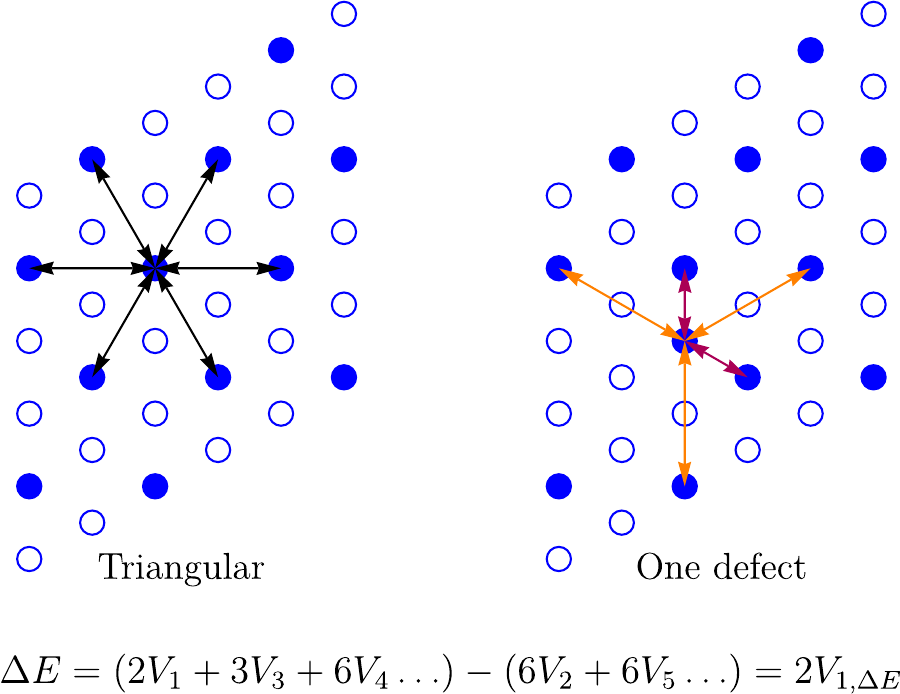}
\caption{\addt{{\bf Schematic of energetic contributions arising from a single defect in the triangular GWC}. The left panel shows the $\sqrt{3} \times \sqrt{3}$ triangular GWC and the underlying original moiré triangular lattice. The right panel corresponds to the configuration where a single charge in this GWC is moved to a vacant nearest neighbor site, creating a ``defect" in the GWC. In both cases some (not all) of the non-zero contributions to the Coulomb energy are highlighted, the expression shown corresponds to the energy difference between the two configurations, which is thus the energy of a single defect $2 V_{1,\Delta E}$ in the effective nearest neighbor model. $V_n$ corresponds to the potential energy arising from the $n^{\rm{th}}$ nearest neighbors.}}
\label{fig:defect_energy}
\end{figure}

\begin{figure}[h]
    \centering
    \includegraphics[width=1\columnwidth]{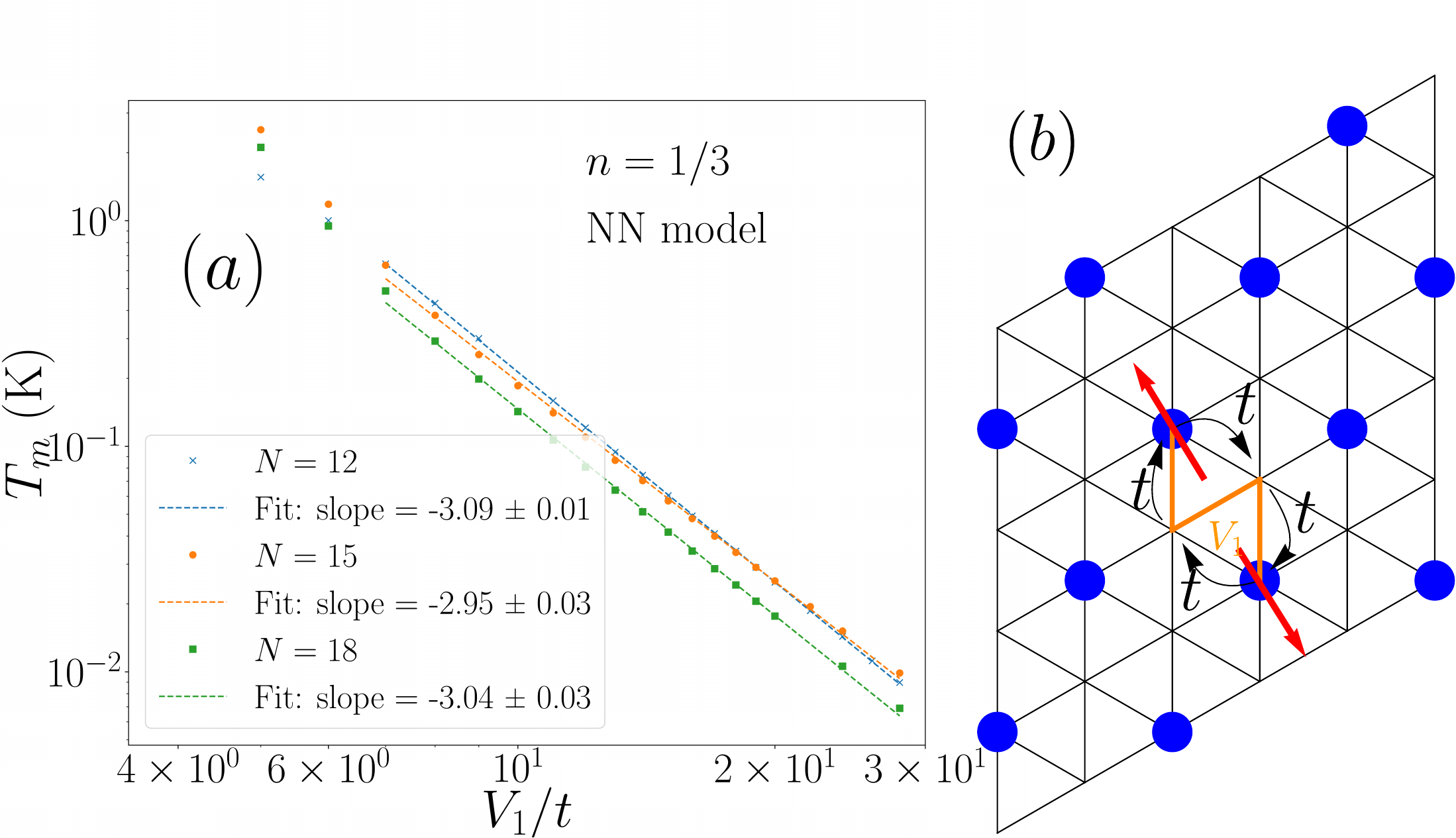}
    \caption{\addt{ {\bf Magnetism in the triangular GWC at $n=1/3$} (a) Magnetic crossover scale $T_m$ as a function of NN interaction $V_1/t$ on a log-log scale for three system sizes $N=12,15,18$ for $n=1/3$.  The overall factor of $t=1.81 $ meV is incorporated in $T_m$ to connect to physical temperature scales. The straight line fit, biased towards larger $V_1/t$, shows the approximate $V_1^{-3}$ dependence. (b) A schematic for the exchange process of two particles with opposite spin involving four hops, that gives rise to the effective magnetic interaction. The orange solid lines show the NN interaction $V_1$ which occurs during the hopping process.}}
    \label{fig:magnetism}
\end{figure}

\noindent
\\
{\bf{Predictions for quantum and thermal melting of GWC }}
\\
We now study the possible quantum mechanical melting of the GWC charge order by controlling the screening environment by tuning $d/a$. 
Smaller $d$ corresponds to a stronger screening, manifesting in a shorter length scale over which the LR Coulomb interaction becomes suppressed by the gate. This suppression of the effective LR Coulomb interaction range with decreasing $d/a$ should eventually drive the system to metallicity despite $U$ being large because of the low density of particles or holes. This is quantitatively established in Fig.~\ref{fig:delta_ED_Tc}a, which shows the value of the order parameter computed on the $N=27$ site cluster (see \addt{Supplementary Note 3}) from ED using the Lanczos method for both $n=1/3$ and $2/3$. (The order parameter corresponds to the Fourier component of the density-density correlation function at the $K$ points.) At large $d/a$, its value is approximately constant, in contrast, on lowering $d/a$, its value rapidly decreases for $d/a \lesssim 1$. We note in passing that given that $a\approx 8$ nm, the $d/a \lesssim 1$ is within experimental range of the hBN dielectric thicknesses \cite{Stepanov2020,Saito2020}.

The dependence of the screened Coulomb interaction from Eq.\eqref{eq:coulomb_long_range} as a function of $d/a$ does not immediately reveal why the order parameter should be essentially constant for large $d/a$. For example, $V_1$ grows (albeit slowly) with $d/a$, see Fig.\ref{fig:delta_ED_Tc}b (red, $V_{1,\text{Coulomb}}$ curve), and \addt{the} tail decays rapidly with $d/a$, see \addt{Supplementary Note 1}. We propose that the origin of the relative independence of the order parameter with $d/a$ stems from the ``cancellations'' of the LR tail which yield an effective NN strength. This NN strength is much smaller than the $V_{1,\text{Coulomb}}$, confirming the expectation we had from the classical specific heat analysis.

\addt{ We verify this hypothesis in two complementary ways, as shown in Fig~\ref{fig:delta_ED_Tc}b. First, at the purely classical level in the LR model (model 2 but with variable $d$), we calculate the energy to create a ``defect", by moving one charge in the \addt{triangular} GWC to a neighboring unoccupied site of the triangular lattice.
Both the GWC and representative defect configuration are shown in Fig.~\ref{fig:defect_energy}. 
For a given $d/a$, the energy difference between the two configurations was computed and defined to be $2 V_{1,\Delta E}$ - this is the energy cost of a defect in the effective NN model. 
$V_{1,\Delta E}/t$ is found to be roughly $6.9$ and is essentially independent of $d/a$ (purple curve in Fig~\ref{fig:delta_ED_Tc}b).  
Fig.~\ref{fig:defect_energy} also graphically depicts the Coulomb energy contributions associated with both configurations. The corresponding expression suggests that their subtraction leads to effective cancellations of the long range tail. It is this cancellation that leads to an energy difference that is (almost) independent of $d/a$.}

In the second method, we vary $V_1$ in the NN model such that its specific heat profile most closely matches that of the LR quantum model.
\addt{We refer to this optimal value of $V_1$ as $V_{1,C,\text{fit}}$}.
For this purpose, we define a cost function (see \addt{Supplementary Note 5}), and we find that this procedure reproduces well both the magnetic bump and charge peak. We find that this value is in the same ballpark, $V_{1,C,\text{fit}}/t \approx 5.9$, and importantly, it is essentially independent of $d/a$ confirming our hypothesis (orange curve in Fig~\ref{fig:delta_ED_Tc}b). \addt{We collectively refer to both these values as $V_{1,\text{eff}}$ with the understanding that it refers to $V_{1,\Delta E}$ in the classical case and $V_{1,C,fit}$ in the quantum case}. We note that both these values of $V_{1,\text{eff}}/t$ \addt{place the material on the insulating side of} the metal-insulator transition (MIT) for $n=1/3$ ($V_{\text{MIT}}/t \approx 4$ \cite{zhou2023quantum}), but \addt{this value of $V_{1,\text{eff}}/t$ is considerably lower} than what has been reported for these moiré materials (e.g. ref \cite{Motruk_2023}). The transition between \addt{metal to insulator} is also expected to be first order~\cite{zhou2023quantum}. This closeness to a phase boundary suggests that local disorder may stabilize pockets of insulating (charge-ordered) and metallic regions. We conjecture that this proximity to the MIT transition \addt{potentially} lies at the origins of the real-space signal variation of what should be a pristine $\sqrt{3}\times\sqrt{3}$ GWC in the STM maps of ref. \cite{feng_li2021imaging}.

The impact of the \addt{renormalization of the NN interaction} is also confirmed by plotting the $T_c$ of the classical and quantum models as a function of $d/a$ in Fig.~\ref{fig:delta_ED_Tc}c. As expected, 
$T_c$ follows $V_{1,\text{eff}}$ from either method rather than $V_{1,\text{Coulomb}}$, which we also check by plotting the ratio of $V_{1,\text{eff}}/T_c$ in the inset. 
(We remark that similar observations for $T_c$ were made for specific values of $d/a$ for the classical model in ref. \cite{mak_xu2020correlated}, \addt{but the origin of this effect was not explained.) Based on our calculations we predict that $T_c$ must not change appreciably for large $d/a$ but at and below $d/a \approx 1-2$ the GWC becomes unstable and melts to give a Fermi liquid.} 

\noindent
\\
\addt{{\bf{Magnetic interactions of GWC for \texorpdfstring{$n=1/3$}{}}}}
\\
\addt{
We conclude by addressing briefly the question of magnetism of GWCs in the moiré TMDs, which remains an active area of research. We present some estimates of the GWC melting (crossover) temperature (which we refer to as $T_m$) for $n=1/3$ as a function of $V_1/t$ in the NN model.
Strictly speaking, there is no true long-range magnetic order in two dimensions at finite temperature, so there is no sharp peak in the specific heat at low temperature, and we associate $T_m$ with the location of the local maximum.}

\addt{
For $V_1/t$ large (and assuming $U \gg V_1$), where the charge order corresponds to the $\sqrt{3} \times \sqrt{3}$ triangular crystal,  the magnetic exchange between two particles with opposite spins is generated (to lowest order) by an exchange process that involves four hops (two hops for each particle) on the underlying triangular lattice, shown in Fig. \ref{fig:magnetism}b.  This gives rise to a magnetic exchange scale that is expected to scale as $t^4/V_1^{3}$. We see evidence of this scale indirectly in $T_m$ which scales as the same power \addt{of $V_1$} (see Fig.~\ref{fig:magnetism}a) . We plot physical estimates for $T_m$ in Kelvin, incorporating the overall scale factor coming from $t$, which clarifies what temperatures should be probed in future experiments. More work is needed to ascertain detailed properties in the context of LR interactions and beyond NN hoppings, along with higher order spin exchange effects.
}


\noindent
\\
\MakeUppercase{\textbf{Discussion}}
\\
In summary, we have studied zero- and finite-temperature properties of GWCs at 
$n=1/3$ and $n=2/3$ filling. We employed a \addt{combination} of numerical techniques to describe an effective extended Hubbard model on the triangular lattice, relevant to moiré materials. Crucially, throughout the manuscript, we related the extended Hubbard model parameters to realistic values expected from the Coulomb interaction and tight-binding estimates of the band structure.

On the theoretical front, our results introduce systematic complementary
procedures for mapping a LR Coulomb model to a NN Hubbard model. Our 
method for matching the specific heat is in the spirit of matching partition 
functions (or density matrices of excited states) that has previously 
been employed in the context of single layer 
graphene~\cite{Schuler_downfolding_2013, Changlani_downfolding_2015, 
Zheng_downfolding_2018}. We showed that when the GWC physics is (largely) controlled by classical effects (i.e., Coulomb interaction dominates; classical and quantum charge order temperature are nearly identical) then the total quantum mechanical LR model can be (approximately) mapped to an NN model by considering the energy of creating a single charge ``defect". 
This renormalization should be strongly dependent on the underlying charge-
ordered GWC itself and it will be interesting to see how this impacts the physics of fillings beyond what has been considered here. We also \addt{remark} that while our renormalization procedure correctly matches the NN and LR charge ordering and magnetic crossover temperatures, this does not necessarily imply that the two models are equivalent in all their aspects; for example, they may differ in the many-body excitation gap \cite{PhysRevLett.115.025701,PhysRevB.84.125120}. 
Understanding the limitations of such effective mappings is something we intend to investigate in future work. More generally, however, we anticipate that this procedure may help in future theoretical studies of GWCs and more generally, moiré systems. 

Through this mapping, we realized that the GWC systems are much closer to the MIT than previous estimates of the NN interaction (based on truncating the LR interaction) would \addt{suggest}.
Our results clarify that the truncation of the LR interaction can stabilize a plethora of charge-ordered states, and must be treated with caution - a sentiment shared with many frustrated magnetic systems where LR tails have been found to be important not only quantitatively but also qualitatively (see for e.g. work in the context of classical spin ice~\cite{Ramirez_1999, Melko_Gingras_2004, Anand_2022}).

Finally, our results explain several outstanding experimental puzzles. First, we showed that despite the kinetic energy being comparable to the charge ordering transition temperature $T_c$, its role is significantly suppressed. However, we demonstrated how the kinetic energy is still important for quantitatively capturing the small asymmetry in the experimentally detected $T_c$ at 
$n=1/3$ and $n=2/3$~\cite{mak_xu2020correlated}. \addt{Next, the conclusion that the effective parameters situate the 
GWC close to the MIT may suggest it is fragile to added perturbations such as disorder, which in turn may be crucial for explaining the origin of the motifs seen in the STM images of ref. \cite{feng_li2021imaging}.} 
\addt{Lastly, we predicted the impact of adjusting the gate-to-sample distance on charge and magnetic ordering temperature scales. The physical scales for the latter are in the ballpark of what is potentially accessible in spin-resolved STM, and when combined with theoretical calculations this could be used to probe the precise microscopic spin ordering patterns of GWCs.}
\\
\\
\noindent
\addt{
\MakeUppercase{\textbf{Methods}}
\\
\\
\noindent
{\bf{Classical Monte Carlo}}
\\
Specific heat calculations for spinless particles (no double occupancy) at fixed density $n=1/3$ were carried out on $N=L \times L$ sized triangular clusters (with periodic boundary conditions in both directions) with classical Monte Carlo, employing the standard Metropolis algorithm. Moves consisted of choosing one occupied site and one unoccupied site at random, and then 
proposing the particle from the former be moved to the latter. One Monte Carlo sweep (MCS) is defined as $N$ moves; measurements were done once every MCS after an initial warm-up stage starting from a completely random configuration. $10^6-10^7$ MCS were used. \\
\\
{\bf{Exact diagonalization, finite temperature Lanczos, and density matrix renormalization group calculations}}
\\
The quantum mechanical models studied in this work utilized a combination of exact diagonalization (ED), finite temperature Lanczos method (FTLM), and density matrix renormalization group (DMRG). 
\\
ED calculations involved either full diagonalization, to compute zero and finite temperature properties, or Lanczos to obtain accurate ground state(s) of finite size clusters; more details on these clusters can be found in Supplementary Note 3. In the case of full diagonalization, expectation values of generic operators $\hat{O}$ were computed in accordance with quantum statistical mechanics, 
\begin{equation}
 \langle \hat{O} \rangle_{\beta} \equiv \frac{\sum_i e^{-\beta E_i} \langle \psi_i | \hat{O} | \psi_i\rangle}{\sum_i e^{-\beta E_i}}
 \label{eq:qstatmech}
\end{equation}
where $E_i$ are exact eigenenergies, $|\psi_i\rangle$ are the corresponding normalized eigenkets, and $\beta$ is the inverse temperature. Translational symmetries (momentum sectors) and $S_z$ conservation were applied to block diagonalize the many-body Hamiltonian and reduce the cost of the computation.
\\
For the ground state Lanczos calculations, $M=100-400$ Krylov vectors were typically used to guarantee convergence on the system sizes studied. In situations where ground state properties were required and the ground state is exactly degenerate (in the same symmetry sector), the algorithm was restarted and the starting vector was explicitly orthogonalized to the previously obtained ground state. Then in accordance with Eq.~\ref{eq:qstatmech} in the limit of $\beta \rightarrow \infty$, the expectation values were averaged over all degenerate states.
\\
FTLM was employed for finite temperature properties where full diagonalization was not possible. The FTLM method utilizes a stochastic representation of Eq.~\ref{eq:qstatmech} and utilizes information from independent Lanczos runs with $R$ starting randomly prepared vectors. Our calculations utilized $M=200$ and $R=10$ per symmetry sector (which for $N=18$ means $R=180$ for every $S_z$ sector) which was found to be sufficient to converge the specific heat up to a scale of $0.1$ K. More details of FTLM and its application to magnetic and fermionic systems can be found in refs.~\cite{Jaklic_Prelovsek, Prelovsek_Bonca, Lee_Sharma_Vafek_Changlani_2023}. Our implementation of ED and FTLM was built on top of the QuSpin library~\cite{quspin} and our own custom codes. 
\\
DMRG is a matrix product state (MPS) variational wavefunction method~\cite{DMRG_White, Stoudenmire_White_review_2012} employed in situations where ED is impractical for obtaining the quantum ground state. The accuracy is controlled by the size of the matrices (bond dimension), and the efficient optimization of the variational parameters entering them using a sweep algorithm. We state the protocols used at various places in the main text and Supplementary Note 2. We carried out DMRG calculations based on the implementation in the TenPy~\cite{tenpy} library.
\\
\\
{\bf{Generation of figures and plots}}
\\
All figures and plots in this paper were generated with a combination of Mathematica~\cite{Mathematica} and Python.
}
\\
\\
\noindent
\addt{
\MakeUppercase{\textbf{Data Availability}}
\\
The data and post-processing scripts used to generate the figures in the present study are available from the first author (A.K.) upon reasonable request.
}
\\
\\
\noindent
\addt{
\MakeUppercase{\textbf{Code Availability}}
\\
The basic scripts and codes used for the simulations reported in this paper are available at elink https://github.com/kaushman1996/quantum-classical-simulation}.
\\
\\
\noindent
\MakeUppercase{\textbf{Acknowledgements}}
\\
We thank V. Dobrosavljevic and V. Elser for useful discussions and for pointing us to the relevant literature and results, and O. Tchernyshyov and S. Sherif for insights from complementary work. \addt{The DMRG and ED simulations were performed
using the TeNPy~\cite{tenpy} and QuSpin \cite{quspin} libraries, respectively. We thank the Research Computing Center (RCC) and the Planck cluster at Florida State University for computational resources. This work also used Bridges-2 at Pittsburgh Supercomputing Center through allocation PHY240324 (Towards predictive modeling of strongly correlated quantum matter) from the Advanced Cyberinfrastructure Coordination Ecosystem: Services and Support (ACCESS) program, which is supported by U.S. National Science Foundation grants $\#$2138259, $\#$2138286, $\#$2138307, $\#$2137603, and $\#$2138296.} We acknowledge support from the National High Magnetic Field Laboratory (NHMFL). The NHMFL is supported by the National Science Foundation through NSF/DMR-2128556 and the state of Florida. A.K. was supported through a Dirac postdoctoral fellowship at NHMFL. C.L. is supported by start-up funds from Florida State University and NHMFL. H.J.C acknowledges funding from National Science Foundation Grant No. DMR 2046570. 
\\
\\
\noindent
\addt{
\MakeUppercase{\textbf{Author contributions}}
\\
C.L. and H.J.C. conceived the initial study, with key inputs from A.K. A.K. and H.J.C. carried out the calculations with additional inputs from C.L. A.K. prepared all figures and H.J.C wrote the paper; all authors contributed to editing the manuscript.
All authors were involved in scientific discussions at all stages.
}
\\
\\
\noindent
\addt{
\MakeUppercase{\textbf{Competing interests}}
\\
The authors declare no competing interests.
}
\\
\\
\noindent
\MakeUppercase{\textbf{References}}

\section*{Supplementary Note 1: Interaction potential and adaptation to finite systems}
\label{sec:app_A}
For two metallic gates placed symmetrically at distance $d/2$ above and below the TMD moiré bilayer, the effective classical potential between two point charges (separated by $\vec{r}$) can be computed with the method of image charges. The result is, 
\begin{equation}
\label{eq:coulomb_app}
V(r \equiv |\vec{r}|) = \frac{e^2}{4\pi \epsilon \epsilon_0 a }\sum^{k=\infty}_{k=-\infty} \frac{(-1)^k}{\sqrt{\Big(\frac{kd}{a} \Big)^2 + \Big( \frac{|\vec{r}|}{a} \Big) ^2}}\,, 
\end{equation}
where $a$ is the moiré lattice constant, $\epsilon_0$ is the permittivity of free space and $\epsilon$ is the dimensionless dielectric constant. Note that we work directly with this expression for 
point charges on the triangular lattice - it is implicitly assumed that the extent of the Wannier function, centered on every moiré triangular lattice site, is much smaller than the distance between the two moiré sites. (See also the discussion in the main text).
In practice, we carry out the summation above by restricting the limits on $k$ from $-K_{max}$ to $K_{max}$ with $K_{max}$ chosen to be $10^{8}$ or $10^{9}$. We use Eq.~\eqref{eq:coulomb_app} for $r<4d$ and for larger $r$ we approximate the potential by its large-distance analytic form,
\begin{equation}
V(r) = \frac{e^2}{4\pi\epsilon\epsilon_0}\sqrt{\frac{8}{rd}} e^{-\frac{\pi r}{d}}\,.
\label{eq:analytic}
\end{equation}

In Supplementary Fig.~\ref{fig:V_r_plots}a we plot $V(r)$, in units of $\frac{e^2}{4\pi \epsilon \epsilon_0 a}$, for various representative values of $d/a$. As expected, for larger $d/a$ the potential is less screened (induced charges on the gates have smaller effect on the interaction between charges) while for smaller $d/a$ the screening is stronger. This is evidenced by a faster decay with $r/a$ for smaller $d/a$. In Supplementary Fig.~\ref{fig:V_r_plots}b we also compare the numerically computed $V(r)$ (Eq.~\eqref{eq:coulomb_app}), for $d/a=10$, with the analytic form (Eq.~\eqref{eq:analytic}) for large $r/a$ -- this confirms the exponential decay of the potential.

Additionally, the overall magnitude of $V_1 \equiv V(r=a)$ increases with increasing $d/a$. In the large $d/a$ limit we are able to approximate the series in Eq.~\eqref{eq:coulomb_app} by the compact expression, 
\begin{equation}
V_1 = \frac{e^2}{4\pi\epsilon \epsilon_0 a} \Big( 1 - \frac{2 \ln 2}{d/a} \Big).
\label{eq:V_1_approx}
\end{equation} 
Both the numerically computed $V_1$ (from Eq.~\eqref{eq:coulomb_app}) and the approximate form (Eq.~\eqref{eq:V_1_approx}) are plotted in Supplementary Fig.~\ref{fig:V_r_plots}c.


All calculations were carried out on finite size systems with periodic boundary conditions (our fundamental unit cell), where one must systematically account for long-range interactions to make meaningful statements about the thermodynamic limit. Conceptually, we imagine all of space to be tiled by translating this unit cell along two lattice vectors. Each charge in the fundamental unit cell then sees another charge in the same cell and its ``images" (not to be confused with the image charges used to derive the screened potential) in the other tiled unit cells. Thus the effective Hamiltonian on finite systems is given by,
\begin{equation}
\label{eq:app_model_ham}
\begin{split}
 H = - \sum_{i<j, \sigma} t_{ij} c^{\dagger}_{i,\sigma} c_{j,\sigma} + \textrm{h.c.} + U \sum_{i} n_{i,\uparrow} n_{i,\downarrow}  \\
+ \sum_{i<j} \tilde{V}_{ij} n_i n_j + 
 \sum_{i} \tilde{V}_{ii} n_i n_i\,,  
\end{split}
\end{equation}
where $t_{ij}$ is assumed to be short-ranged (only nearest neighbor in this work) and $\tilde{V}_{ij}$ and $\tilde{V}_{ii}$ are effective interaction potentials that account for the intra and inter-unit cell contributions given by the expressions,
\begin{eqnarray}
    \tilde{V}_{ij} &=& \sum_{n_1,n_2 = -\infty}^{\infty} V(\vec{r}_{ij}+n_1 \vec{L}_1 + n_2 \vec{L}_2)\,, \\
    \tilde{V}_{ii} &=& \sum_{\substack{n_1,n_2 = -\infty \\ (n_1,n_2) \neq (0,0)}}^{\infty} V(n_1 \vec{L}_1 + n_2 \vec{L}_2) \,,   
\end{eqnarray}
where $\vec{L}_1$ and $\vec{L}_2$ are inter unit cell translation vectors. 
In practice, these sums cannot be carried out exactly, and must be approximated (to very high accuracy) by truncating the limits on $n_1$ and $n_2$ to $ \pm N_{max}$. For all $d/a$ and system sizes considered in the paper, we find that it is sufficient to set $N_{max} = 100$. The rapid convergence with $N_{max}$ is a direct result of the exponential decay - it is known that when the interactions are not screened, methods such as Ewald summation must be applied to obtain the effective potential~\cite{deLeeuw_1980,Melko_Gingras_2004, Anand_2022}.

\begin{figure}
\includegraphics[width=1\columnwidth]{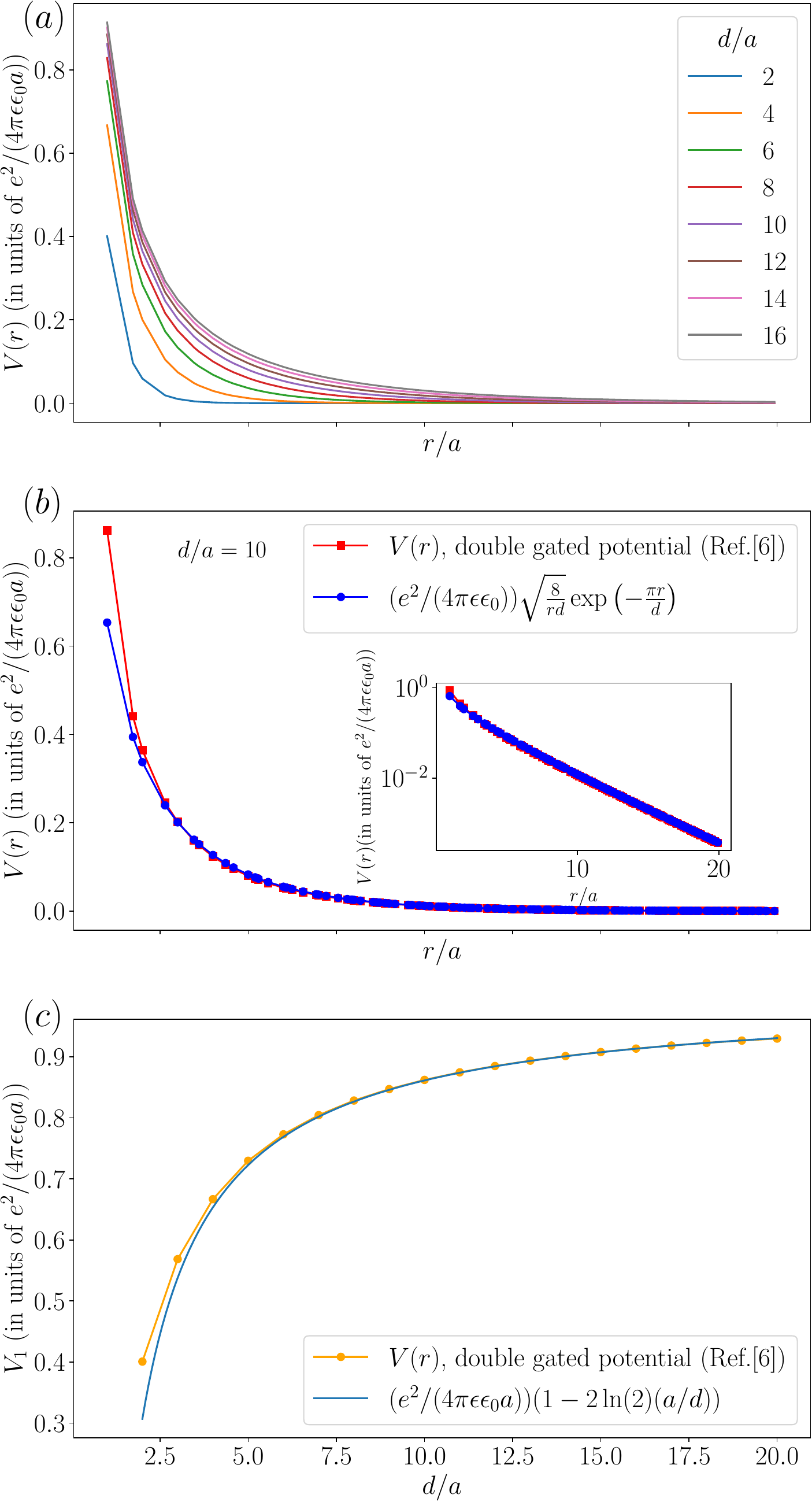}
\caption{(a) Interaction potential $V(r)$, as in Eq.~\eqref{eq:coulomb_app}, vs $r/a$ for various representative values of $d/a$. (b) $V(r)$ vs $r/a$ for the specific case of $d/a=10$ showing comparison with the analytic asymptotic form, Eq.~\eqref{eq:analytic}, for large $r/a$. The inset shows the excellent agreement between the curves on a log scale. (c) Nearest neighbor interaction $V_1 \equiv V(r=a)$ as a function of $d/a$. The asymptotic expression for large $d/a$, Eq.~\eqref{eq:V_1_approx}, is also plotted.}
\label{fig:V_r_plots}
\end{figure}

\section*{Supplementary Note 2: Density Matrix Renormalization Group (DMRG) results on cylinders}
\label{sec:app_B}
In the main text we discussed results based on ground state DMRG calculations carried on the $6 \times 6$ torus geometry. 
Since the kinetic energy (quantum) term is small compared to the Coulomb energy, and there are many competing classical configurations, some dependence on the finite system shape and size is expected. 

To build confidence in our conclusions based on the $6 \times 6$ torus shown in the main text, we have carried out additional DMRG calculations on cylindrical geometries of width 6, restricting the calculation to the $S_z = 0$ sector. In Supplementary Fig.~\ref{fig:cylinderical_boundary_result} we show results for the local charge density $\langle n_i \rangle$ (on every site $i$) computed in the ground state of three finite-range models that were obtained by truncating the full long-range (LR) model i.e. model 3 of Table 1 in the main text. These models correspond to nearest-neighbor (NN) (Supplementary Fig.~\ref{fig:cylinderical_boundary_result}a), next-nearest neighbor (NNN) (Supplementary Fig.~\ref{fig:cylinderical_boundary_result}b) and next-next-nearest neighbor (NNNN) interactions (Supplementary Fig.~\ref{fig:cylinderical_boundary_result}c). Note that model 3 when truncated up to NNNN interactions is exactly equivalent to model 4. In all three cases the hopping is taken to be NN. The diameter of the blue circles in the figure is proportionate to $\langle n_i \rangle$.

For the NN (i.e. $t-V_1$) model we see a clear $\sqrt{3} \times \sqrt{3}$ triangular crystal. The blue circles correspond to a charge density close to (but less than) 1, suggesting some melting of the classical generalized Wigner crystal (GWC) order due to quantum effects.
\begin{figure}
    \centering
    \includegraphics[width=1\linewidth]
    {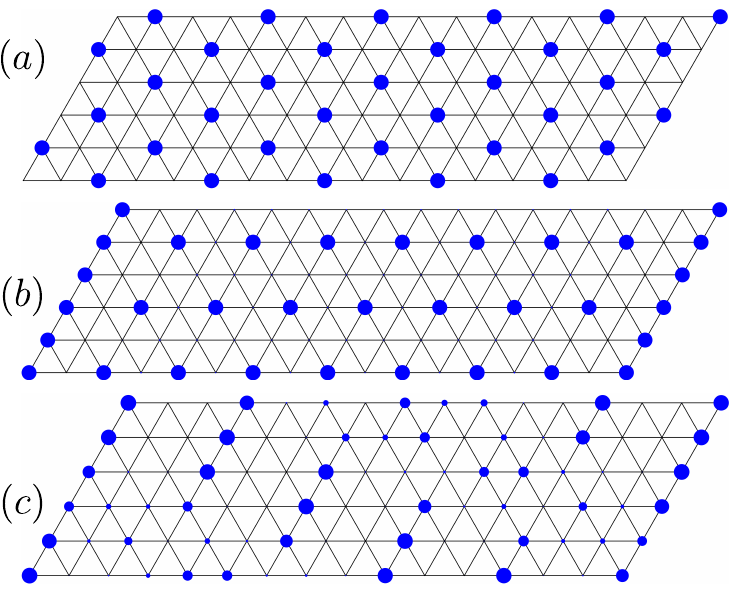}
    \caption{Ground state local charge density $\langle n_i \rangle$, computed with DMRG, on every site $i$ of a cylinder, for three models obtained from the truncating the LR model (model 3 of Table 1 of the main text). These models correspond to interactions up to (a) NN, (b) NNN and (c) NNNN. The sizes of the blue circles shown are proportionate to $\langle n_i \rangle$. The DMRG simulations employed a maximum bond dimension of 8000 with a truncation error of $O(10^{-6}).$ The total energy during the sweeping process was converged to $O(10^{-7})t$.}
    \label{fig:cylinderical_boundary_result}
\end{figure}
For the NNN (i.e. $t-V_1-V_2$) model we observe a $2\times2$ Wigner crystal in the bulk with delocalized charges -- this is the pinball liquid phase. To observe this pattern we minimized boundary effects by adjusting the overall charge density. Specifically, we simulated charge densities in the vicinity of (and slightly higher than) 1/3, the charges in excess of the 1/3 density migrated to the open boundary edge - this is the system's mechanism for minimizing the Coulomb cost. The bulk of the cylinder was left with an average density of 1/3, which we take to be representative of the thermodynamic limit. We have also shown only the result from a representative cluster that has dimensions $17 \times 6$ lattice constants - this accommodates a robust charge order throughout the entire cylinder (except the edges). On systems that had an even number of lattice constants in the horizontal direction (not shown), the $2\times2$ charge orders propagate from the left and right edges, creating domains that results in the appearance of a liquid-like region at the domain wall (i.e. in the central region of the cylinder). 

\begin{figure}[h]
    \centering
    \includegraphics[width=1\linewidth]{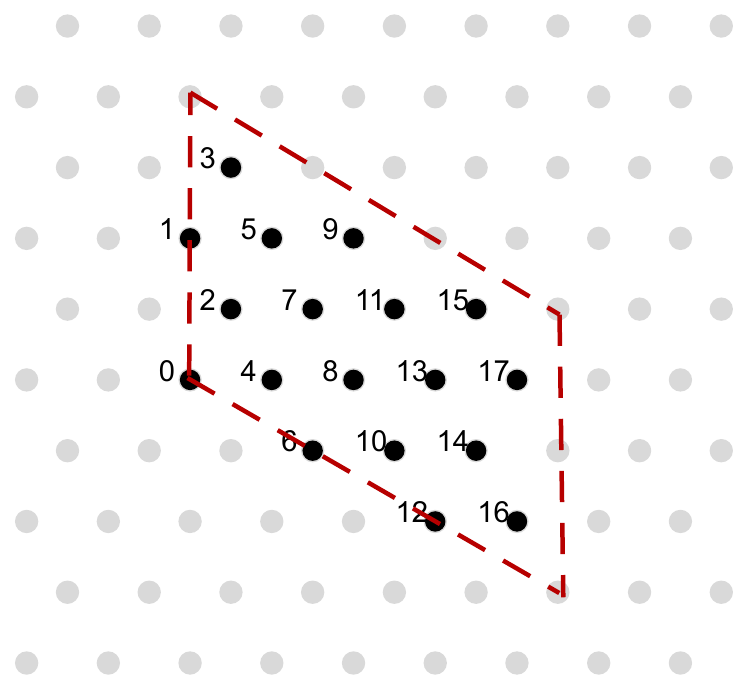}
     \includegraphics[width=1\linewidth]{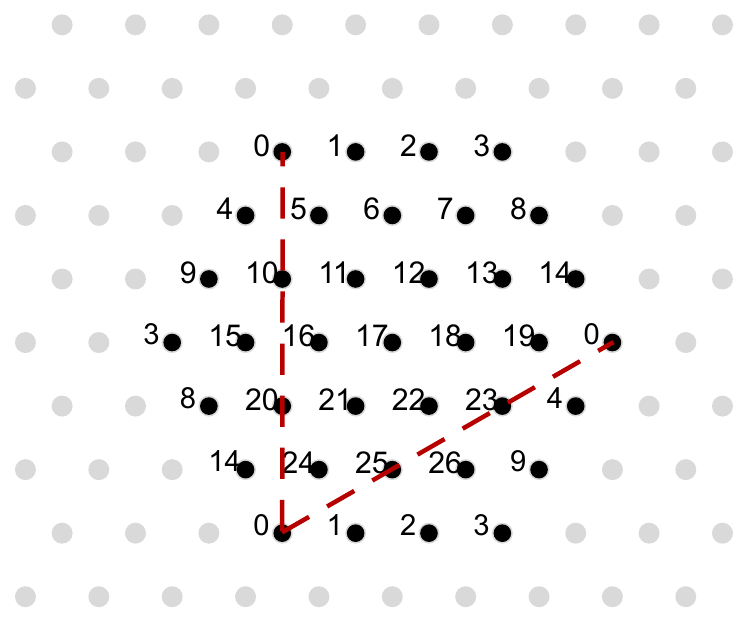}
    \caption{$N=18$ and $N=27$ clusters with periodic boundary conditions, utilized for exact diagonalization calculations in this work.}
    \label{fig:N_18_27}
\end{figure}

The NNNN (i.e. $t-V_1-V_2-V_3$) model shows the formation of stripes, however, unlike the case of the $6\times 6$ torus these do not extend throughout the width of the cylinder (at least for long cylinders). Instead the system tends to form only short stripes, qualitatively consistent with what was seen in classical Monte Carlo at low temperatures, for systems bigger than $6 \times 6$. It is also possible that energetic reduction can be achieved from enhanced quantum effects associated with shorter stripes. Additionally, the parameters of the $t-V_1-V_2-V_3$ model place the system near a classical phase boundary suggesting sensitivity to boundary effects. Thus, further investigation is required to probe the true nature of the bulk ground state in the thermodynamic limit.

\begin{figure}
	\includegraphics[width=\linewidth]{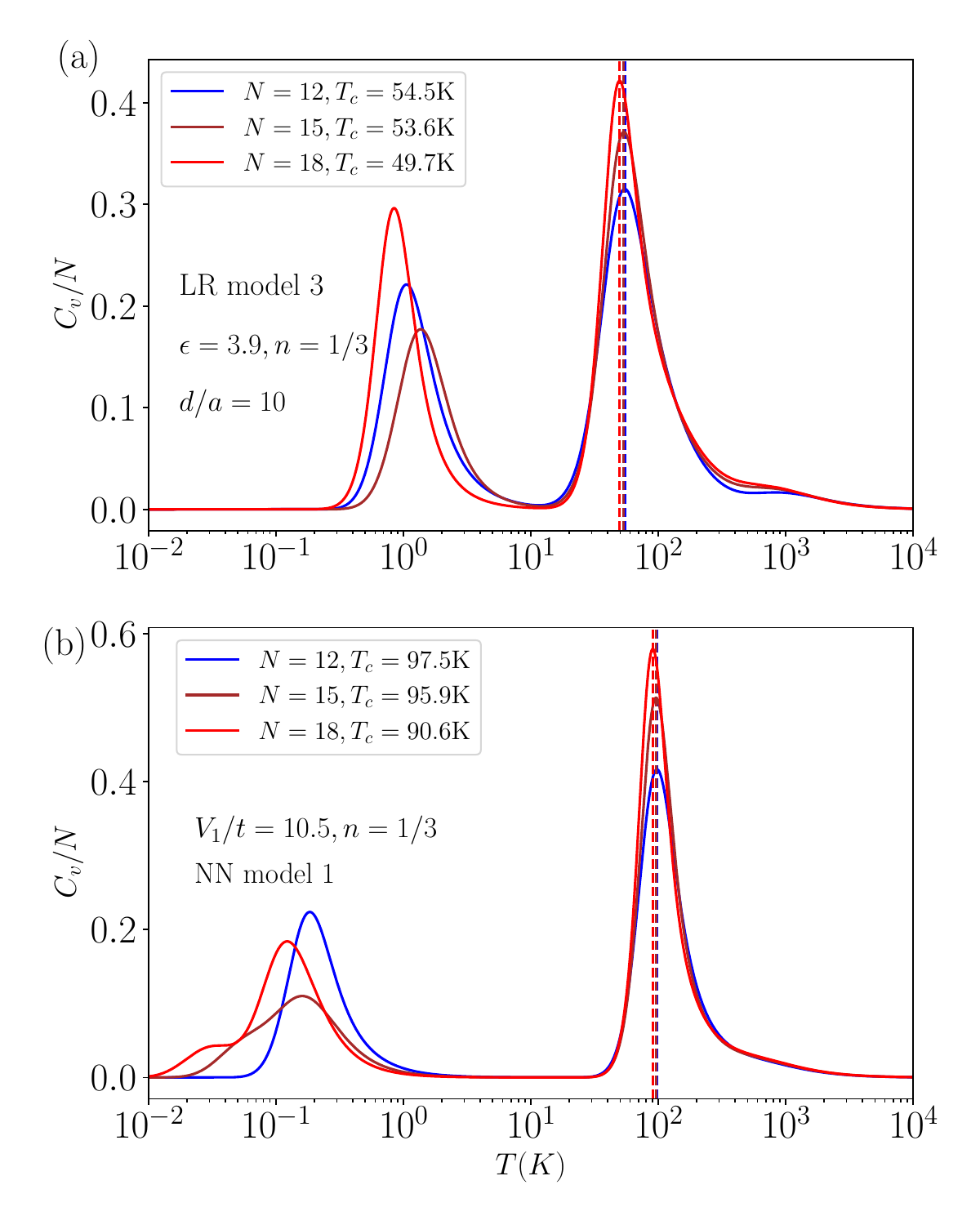}
	\caption{Specific heat per site ($C_v/N$) as a function of temperature ($T$), computed with full diagonalization for $n=1/3$, (a) for the LR model (model 3) and (b) NN model (model 1) for three system sizes $N=12,15,18$ with periodic boundary conditions. Some of the parameter values in these models are indicated on the figures, for more details refer to Table 1 of the main text.}
 \label{fig:cv_quantum_finite_size_appendix}
\end{figure}

\section*{Supplementary Note 3: Cluster geometries used for exact diagonalization}
\label{sec:app_C}
In the main text we performed exact diagonalization (ED) calculations (either full diagonalization or Lanczos) to compute ground state and finite temperature properties. The $N=12$ and $N=15$ clusters utilized were considered in previous work~\cite{Lee_Sharma_Vafek_Changlani_2023}. In Supplementary Fig.~\ref{fig:N_18_27} we show the $N=18$ and $N=27$ clusters with periodic boundary conditions that were utilized for ED calculations in this work.


\section*{Supplementary Note 4: Finite size effects on features of the specific heat}
\label{sec:app_D}
In the main text we presented results for the specific heat of classical models ($t=0$, $U \rightarrow \infty$) for a wide range of system sizes. Despite obvious limitations of finite-size effects, meaningful trends could be extracted even from the smallest system sizes that were studied. For example, the location of the charge peak, corresponding to the melting of the triangular GWC, moves to lower temperature on increasing the system size, and for $N=27$ and $36 \times 36$ these were within about $10\%$ of one another. 

We show that similar conclusions hold at the quantum level by plotting the specific heat per site ($C_v/N$) for the LR model (model 3, see Supplementary Fig.~\ref{fig:cv_quantum_finite_size_appendix}a) and the NN model (model 1, see Supplementary Fig.~\ref{fig:cv_quantum_finite_size_appendix}b)
for $N=12,15$ and $18$ site systems with periodic boundary conditions. 
Each finite size cluster accommodates the $\sqrt{3}\times\sqrt{3}$ triangular GWC ground state. 

On increasing system size the high temperature feature sharpens (narrows in extent) while also achieving a higher maximum of specific heat. This is consistent with a peak that signals a phase transition associated with the melting of the GWC. The location of the peak ($T_c$) moves to lower temperature on increasing system size. 
The low temperature feature (bump) is associated with the destruction of short range magnetic order. While the rough scale is consistent between system sizes, a quantitative explanation of the movement of the low temperature features with system size (and shape) is likely more involved and requires a better understanding of the effective magnetic Hamiltonian that governs the physics at low energy scales. 

\section*{Supplementary Note 5: Determination of $V_{1,C,fit}$ for the effective nearest neighbor model from the specific heat of the long-range model}
\label{sec:app_E}
In the main text  we showed that certain properties of the LR model can be mimicked by an effective NN model. This was particularly important when trying to understand why the GWC melting temperature did not change significantly with $d/a$.

Here we show a representative example for how $V_{1,C,fit}$ was obtained using ED. First, the specific heat was computed for the LR model (model 3 but with a variable $d$) for a given $d/a$. Then keeping $t$ unchanged, the specific heat was computed for the NN model for a grid of values of $V_{1}$, chosen in steps of $0.1 t$. The optimal $V_{1,C,fit}$ was chosen to be the one which minimized the error between the $T_c$ of the two specific heat curves. 
Representative examples of such optimizations have been presented in Supplementary Fig.~\ref{fig:Cv_overlaid}. 

\begin{figure}[h]
\includegraphics[width=\columnwidth]{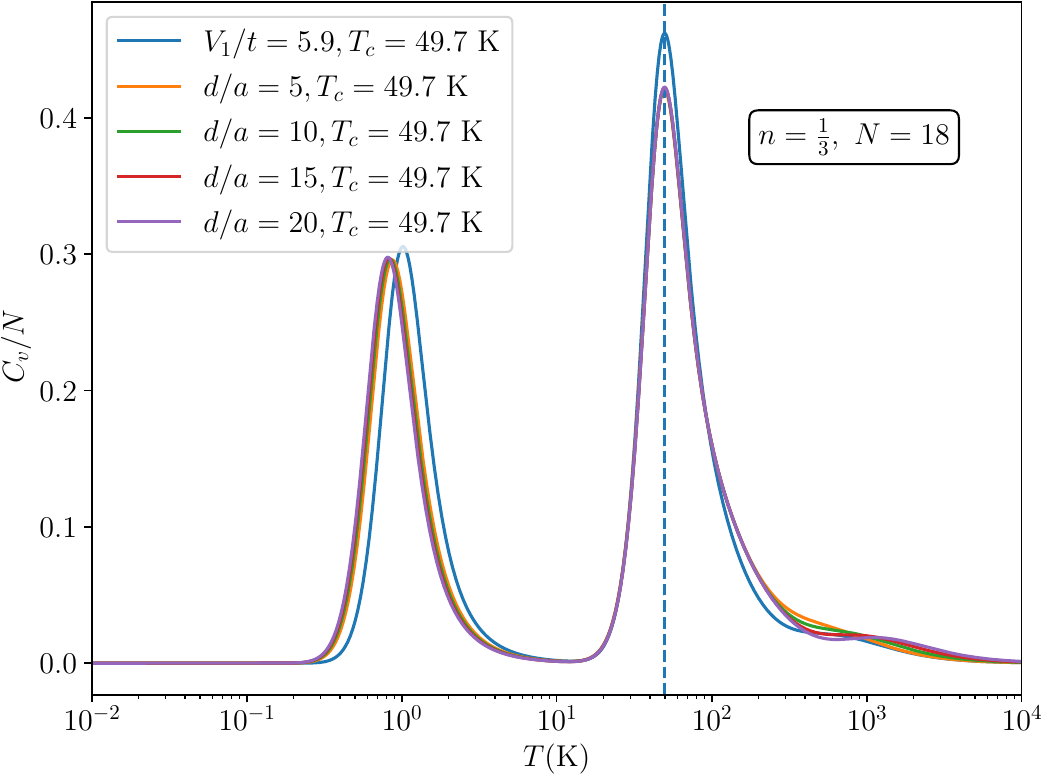}
\caption{Specific heat per site ($C_v/N$) for $n=1/3$, as a function of temperature ($T$), for the LR model with $\epsilon =3.9$, for a set of $d/a$ values $(5,10,15,20)$ (model 3) plotted together with that of the NN model with $V_{1}/t=5.9$.}
\label{fig:Cv_overlaid}
\end{figure}
\end{document}